\documentclass[11pt]{article}  

\RequirePackage{graphicx}
\RequirePackage{amsmath,amssymb}
\RequirePackage{epsfig}
\RequirePackage[dvips]{color}

\RequirePackage{latexsym}

\newcommand{\be}{\begin{equation}}
\newcommand{\ee}{\end{equation}}
\newcommand{\bes}{\begin{subequations}}
\newcommand{\ees}{\end{subequations}}
\newcommand{\bea}{\begin{eqnarray}}
\newcommand{\eea}{\end{eqnarray}}
\newcommand{\bear}{\begin{equation}\begin{array}}
\newcommand{\eear}[1]{\end{array}\label{#1}\end{equation}}
\newcommand{\bu}{$\bullet$\ }

\newcommand{\lb}{\linebreak[4]}

\newcommand{\fr}[2]{\dfrac{{ #1}}{{ #2}}}

\newcommand{\fn}[1]{\footnote{{#1}}}
\renewcommand{\le}{\leqslant}

\def\vep{{\varepsilon}}
\newcommand{\epe}{\mbox{$e^+e^-\,$}}
\newcommand{\ggam}{\mbox{$\gamma\gamma\,$}}

\newcommand{\cl}{\centerline}

\begin{document}

\title{Triple Higgs coupling in the most general 2HDM at SM-like scenario}

{\author{ Ginzburg I.F.\thanks{e-mail: ginzburg@math.nsc.ru} $^{,1,2}$}
\maketitle
{\it $^1$ Sobolev Inst. of Mathematics, av. Koptyug, 4, Novosibirsk, 630090, Russia,}

 {\it$^2$ Novosibirsk State University,  Pirogova str., 2, Novosibirsk, 630090, Russia}}

\maketitle
\begin{abstract}
{ We consider the  triple Higgs coupling for $h(125)$ Higgs boson
within the most general 2HDM. At  moderate values of parameters of
model, allowing by modern data, noticeable deviation of this
coupling from   its SM value is improbable. This deviation can be
sizable only if some measurable parameters of the model are
exotic.

%
}
\end{abstract}

\flushbottom


\section{Introduction}

The recent discovery of a Higgs boson with $M\approx 125$~GeV at
the LHC \cite{125Higgs-1}-\cite{125Higgs-4} suggests that the
spontaneous electroweak symmetry breaking is most probably brought
up by the Higgs mechanism. The simplest realization of the Higgs
mechanism introduces a single scalar isodoublet $\phi$ with the
Higgs potential\\ \cl{$V_H=-m^2(\phi^\dagger \phi)/2
+\lambda(\phi^\dagger \phi)^2/2$.} This model is usually called
"the Standard Model" (SM).  The mentioned  data do not rule out
the possibility of realization of {\it beyond SM models} (BSM)
which include both neutral Higgs scalars $h_a$ (generally without
definite CP parity)  and charged Higgs scalars $H^{\pm}_b$  with
masses $M_a$ and $M_{b\pm}$ respectively.

In the  discussion that follows we use the   relative couplings
for each neutral Higgs boson $h_a$ (for the case with single
charged Higgs boson $H^\pm$): \bear{c}
\chi^P_{a}=\fr{g^P_a}{g^P_{\rm SM} }\;\; \left[ P=V\,(W,Z),\;
q=t,b,...,\; \ell=\tau,...\right] \,,
\\[3mm]
\chi^\pm_a=\fr{g(H^+H^-h_a)}{2M_\pm^2/v},\quad  \chi_{a}^{H^+ W^-} = \dfrac{g(H^+ W^- h_a)}{M_W/v}\,,\\[3mm]
\chi_{abc}=\fr{g(h_ah_bh_c)}{g(hhh)_{SM}}\,.
\eear{relcoupl}
 The quantities $\chi^P_{a}$ are the
ratios of the couplings of  $h_a$   with the fundamental particles $P$
to the corresponding  couplings for the would be SM Higgs boson with $M_h=M_a$.  The   other relative couplings describe interaction of $h_a$ with charged Higgs boson.
Couplings $\chi^V_a$ and $\chi^{\pm }_a$ are real due to Hermiticity of Lagrangian, while other couplings  are generally complex.

We  omit the adjective "relative" below.

\subsection{SM-like scenario}

Current data allow us to suggest
 that Nature realizes the  {\it SM-like scenario}\fn{ The term {\it SM-like scenario} was introduced in \cite{GKO}, the term {\it alignment limit} was introduced recently for this very situation, see e.g. \cite{align},  the {\it decoupling limit} is the particular case of this scenario.}:
{\it The observed particle  with mass $M\approx 125$~GeV is a Higgs boson,  we call it  $h_1$. It interacts with the gauge bosons and $t$-quarks with coupling strengths that are close to those predicted by the SM
within experimental accuracy (see e.g. \cite{125_2HDM-1}-\cite{125_2HDM-3})}. In particular, for coupling with the gauge bosons
 \be
\vep_V=\left|1-(\chi^V_1)^2\right|\ll 1\,.\label{SMlike}
 \ee
 In estimates we will have in mind $\vep_V\le 0.1$.
\subsection{Two Higgs doublet Model (2HDM)}

The 2HDM presents the simplest extension of the standard Higgs model \cite{TDLee}. It offers a number of phenomenological scenarios with different physical content in different regions of the model parameter space, such as a natural mechanism for spontaneous CP violation, etc. \cite{TDLee}-\cite{Branco2HDM} For example, the  Higgs sector of the MSSM is a particular case of 2HDM. Some variants of 2HDM have interesting cosmological consequences \cite{GIK09-1},\cite{GIK09-2}.

In the most general 2HDM    the couplings \eqref{relcoupl} obey the following sum rules  \cite{gunion-haber-wudka}-\cite{GKan}:
 \bes\label{SRV}\bea
\sum\limits_a (\chi^V_{a})^2=1,\\
 |\chi_{a}^V|^2+| \chi_{a}^{H^\pm W^\mp}|^2=1
,\\ \sum\limits_a (\chi^f_a)^2\!=\!1.
 \eea\ees

We have constructed in  \cite{GKan}  the minimal complete set of measurable quantities ("observables") which  determine  all parameters of the  2HDM.  This set contains
\bear{l}
\mbox{\it v.e.v. of Higgs field  $v=246$~GeV},\\ \mbox{\it masses of  Higgs bosons } M_a,\; M_\pm\;\;\;(a=1,\,2,\,3),\\ \mbox{\it two out of three couplings } \chi_a^V,\\
\mbox{\it 3 couplings }H^+H^-h_a \;\; \mbox{\it(quantities }\; \chi^\pm_a\;Eq. \eqref{relcoupl}),\\  \mbox{\it quartic coupling } g(H^+H^-H^+H^-).
\eear{setpar}
 In the most general 2HDM,  these observables are independent of each other. In some particular variants of 2HDM,  additional relations between these parameters may appear (for example, in the CP conserving case  we have  $\chi^V_3=0$, $\chi_3^\pm=0$).

\subsection{Limitations for parameters}

The values of parameters $\lambda_a$ of 2HDM (and -- therefore -- mentioned basic parameters)  obey two groups of constraints  (see e.g. \cite{decoupling}, \cite{Branco2HDM}).

{\bf Positivity constraints} are conditions for the stability of Higgs potential at large quasi-classical values of fields.  They do not  restrict  parameters from above.

{\bf Perturbativity  (and unitarity) constraints} make it possible to use the first non-vanishing approximation of perturbation theory for description of physical phenomena with reasonable accuracy -- {\it perturbative description}. (This is a tree approximation for most of phenomena and a one-loop approximation for the phenomena which are absent at tree level, e.g. decays $h\to\ggam$, $h\to Z\gamma$, $h\to gg$). The starting point in  obtaining of these constraints is the observation that the effective parameter of perturbative expansion is not $\lambda_i$ ($i=1,2,...7)$ but $\lambda_i/\Delta $ with $\Delta =8\pi$ or $4\pi$.  The perturbativity condition is written usually in the form $|\lambda_i|<\Delta$.

At $|\lambda_i|\approx \Delta$  perturbative description of physical  phenomena is incorrect even at low energies. In particular,  the  equations, expressing masses and couplings via parameters of Lagrangian, become invalid. Good  example provide   the one-loop radiative corrections (RC) to the triple Higgs coupling \cite{RCShinya}-\cite{ArhribRC}. In the   SM-like scenario these RC  reach $150\div 200\% $ at $|\lambda_i|\approx\Delta$. (The ref.~\cite{ArhribRC}  presents example with clear details. Authors consider  Inert Doublet Model, i.e. 2HDM with exact $Z_2$ symmetry in the SM-like case, at $\lambda_4=\lambda_5=0$ and $\lambda_1=\lambda_{SM}$. The one-loop corrections to the $g(h_1h_1h_1)$ are described by single parameter $\lambda_3$, they reach $180\%$ at $|\lambda_3|\approx\Delta$.)

The first non-vanishing approximation of perturbation theory describes physical phenomena with  relative inaccuracy $k$ only at
\be
|\lambda_i|<k\Delta\,\qquad (k<1)\,.\label{pertlim}
\ee
In particular, in the region of parameters, provided accuracy of standard description in 30\% one should have $k=0.3$. In this region of parameters the value of RC, discussed in \cite{RCShinya}-\cite{ArhribRC}, does not exceed 20\%.

Below we will have in mind this very limitation with $k\approx 0.3$.

{\bf The realization of the SM-like scenario imposes additional restrictions on the parameters}.
Because of sum rules \eqref{SRV}, in the SM-like scenario the couplings of  other neutrals $h_a$ with gauge bosons $\chi^V_a$ are small. Besides, the absolute value of non-diagonal coupling  with EW gauge bosons for the observed Higgs boson $\chi^{W^\pm H^\mp}_1$ is small\fn{The calculations of $H^-\to W^-h_1$ decay at LHC in \cite{WHh1-1},\cite{WHh1-2} are made in the CP-conserving 2HDM  and with  not very small $\vep_V$.}, while similar couplings for the  other neutrals $\chi^{W^\pm H^\mp}_{2,3}$ are close to their maximal possible values:
  \bea
\boxed{ (3a)}\;\Rightarrow\;\;&  |\chi^V_a|^2<\vep_V\ll 1\,, \quad a=2,\,3\,.\label{chVa}\\
\boxed{ (3b)}\;\Rightarrow\;\;& |\chi^{H^\pm W^\mp}_1|^2\sim \vep_V\ll 1\,;\;  \;\;|\chi^{H^\pm W^\mp}_{2,3}|^2\approx 1\,.\label{chWa}
\eea

 {\bf In the  SM-like scenario the perturbativity constraints lead to additional restrictions.} In particular, according to  Eq.~(23) from Ref.~\cite{GKan}, the perturbativity constraint \eqref{pertlim} imposes the limitation on the coupling of $h_1$ to charged Higgs bosons:
\be
|\chi_1^\pm|<1 \;\;\mbox{at}\;\; M_\pm>500~\mbox{GeV}.\label{limchcoupl}
\ee
It means that  the heavy charged Higgs boson gives only small contribution to the two-photon width of the observed Higgs boson $h_1$.

Next, we consider heavy neutral Higgs bosons $h_a$ ($a=2,\,3$) in the SM-like scenario.  The couplings $\chi_a^V$  are small (see  \eqref{chVa}), while Eq.~(23) from Ref.~\cite{GKan} allows to have big values of $\chi_a^\pm$
 ($\lesssim 1/\sqrt{\vep_V}$).
Therefore, the two-photon width of the boson $h_a$
is strongly different from the similar width calculated for the would-be SM Higgs boson with the  mass $M_a$.

\section{Triple Higgs vertex}

The observation  of $hh$ production and the extraction of the triple Higgs vertex $g(hhh)$ from the future experiments
is scheduled at the LHC and other colliders. This is a necessary step in the verification of the Higgs mechanism. Hopefully, these observations will allow us to see the effects of BSM\fn{In the models, containing additional heavy Higgs bosons $h_a$ with $M_a>2M_1$, the {\bf resonant} $h_1h_1$ production like $pp\to (h_2\to h_1h_1)+...$ becomes possible. In this paper we  discuss only non-resonant $h_1h_1$ production, without intermediate $h_a\neq h_1$.}.

The studies of triple Higgs coupling have long history, for recent reviews see e.g. \cite{revhhhmeas}.  There are two major parts. The first one is whether it is possible or not to observe $hh$ production, caused by $hhh$ vertex. The second one is whether it is possible to use these observations for extraction of New Physics effect beyond SM.

The accuracy  in the  extraction of a triple  Higgs vertex $g(hhh)$ from the future data cannot be high,  since in each case corresponding experiments deal with interference of two channels  -- an independent production of two Higgses and production of Higgses via $hhh$ vertex. This interference is mainly destructive \cite{hhhinterf}. For example, for 100~TeV hadron collider with total luminosity 3/{\it ab} one can hope to reach accuracy  of 40\% in the extraction of this vertex from future data \cite{hhhestim};  at ILC the accuracy in the extraction of $g(hhh)$ will be better than 80\% only after 10 years of operation \cite{ILCprogr}.   Therefore, the effects of New Physics will be distinguishable in the  data of $g(hhh)$ in the realistic future only if the deviation of this coupling from its SM value is
high enough,
\be
\left|\chi_{111}-1\right|\gtrsim 1.\label{criterium}
\ee

One of the approaches in the description of the SM  violations is to add in the SM Lagrangian  terms with anomalous interactions of Higgs boson. It was found for many reasonable benchmark points that these anomalous interactions are difficult for observation \cite{anomk}.

The other approach is to consider some special form of BSM. The review of the whole variety of possible BSM models is
beyond our scope. We limit ourself considering 2HDM
in its most general form.

The potential of such $g(h_1h_1h_1)$ observation was studied for some benchmark points of parameters  of 2HDM mainly in the case of CP conservation and with moderate values of parameters  \cite{hhhviol1}, for the  MSSM with CP-conservation  \cite{hhhMSSMn-1} or with violated CP \cite{hhhMSSMn-2} mainly beyond SM-like scenario. The case of SM-like scenario with similar limitations was considered in \cite{hhhviol}.

\subsection{\bf Triple Higgs coupling via observables}\label{sechhh}

The transition from neutral components of basic fields
$\phi_{1,2}$ to the neutral Higgs bosons $h_a$ is described by
some mixing matrix. The equation for triple Higgs coupling in the
most general 2HDM via parameters of Lagrangian and elements of
this mixing matrix is obtained simply (see e.g. Eq. (25) of Ref.
\cite{GKan}). The expression of this coupling in terms of the
introduced observables  \eqref{setpar} was obtained in Eq. (36) of
Ref. \cite{GKan}. We transform it to the following form:
\bear{c}
g(h_1h_1h_1)=(M_1^2/v)\,\chi_{111};\\[2mm]
\chi_{111}\!=\!
\chi^V_1\left[1\!+\!\left(1\!-\!(\chi^V_1)^2\right)R\right],\quad
R\!=\!\sum\limits_{i=1}^3R_i\,;\\[2mm]
R_1\!=\!\fr{2M_\pm^2}{M_1^2}\left(\chi^V_1\chi^\pm_1\!-\!1\right),\;
R_2\!=\!1\!
+\!2\sum\limits_{b=1}^3 \fr{M_b^2}{M_1^2}(\chi^V_b)^2,
\\[3mm]
\!R_3\!=\!\fr{2M_\pm^2}{M_1^2}\!\!\left[\!\!\sum\limits_{b=2}^3\chi^V_b\chi^\pm_b\!\!+\!
Re\!\left(\!\fr{\chi^{H^-W^+}_1\!\!}{\chi^V_1}\!\!\sum\limits_{b=1}^3
\chi^{H^+W^-}_b\!\!\!\chi^\pm_b\right)\!\!\right]
\eear{triple1-c}

\subsection{\bf Triple Higgs coupling in SM-like scenario}

 With  estimates \eqref{chVa}, \eqref{chWa} we have
\be
\chi_{111}\approx
[1+(R-1/2)\vep_V]\,.
\label{triple1est}
\ee

We see that  at  moderate values of parameters, the relative coupling $\chi_{111}$ is close to 1, and it is difficult to expect a sizable effect\fn{For the particular CP conserving case and with moderate values of parameters such conclusion was obtained in \cite{hhhviol}, \cite{hhhviol1} (see also \cite{hhhMSSMn-1}, \cite{hhhMSSMn-2} for  the CP conserving MSSM).}.

 Nevertheless, it is interesting to consider the {\it special exotic  values of model parameters} that provide sizable deviations of triple Higgs coupling from its SM value, i.e. $|\chi_{111}-1|\gtrsim 1$.   We consider the  effect of different terms $R_i$, entering $R$.

 The term  $R_1$ can give $|\chi_{111}-1|\gtrsim 1$ if the charged Higgs boson $H^\pm$ is heavy enough  and  the
coupling $\chi_1^\pm$ of the observed Higgs boson with $H^\pm$ deviates substantially from the  value $\chi_1^\pm\approx 1$. In view of Eq.~\eqref{limchcoupl}, it can happen if this coupling is either very small or negative.

The term $R_2\!\sim\! 3\!+\!\vep_V (M_b^2/M_1^2)$ can give\lb $|\chi_{111}-1|\gtrsim 1$  only if at least one of other Higgs bosons $h_{2,3}$ is heavy enough, $M_{2,3}>1$~TeV. Direct discovery of such Higgs bosons seems to be a difficult task.  Therefore  the value of $g(h_1h_1h_1)$ might become an important source of knowledge about such heavy neutrals for a long time.

 The term $R_3$ contains  small factors $\chi_{2,3}^V$,  $\chi_1^{H^+W^-}$  and factors $\chi_{2,3}^\pm$ which can be large (up to $1/\sqrt{\vep_V}$). The term $R_3$ can be not small if $H^\pm$ is heavy.

Certainly, the real range of possible values of discussed parameters is restricted  by  other observations. Better estimates are possible only after measuring of $\vep_V$ with reasonable accuracy. In particular, at $\vep_V\ll 0.1$ we cannot expect sizable effects in the triple Higgs vertex.

\section{Summary}

\bu \   Measuring  $hh$ production at various colliders is a necessary step in the verification of Higgs mechanism of EWSB.
Within the  SM-like scenario in the 2HDM, these measurements can give information about the New Physics
beyond SM only at exotic values of the parameters listed above.  The enlargement of the field of parameters of 2HDM at the transition from CP conserved softly $Z_2$ broken potential to the most general case gives no new essential opportunities in the deviation  of triple Higgs coupling from its SM value.

In our conclusions we limit ourself perturbative limitations in the form \eqref{pertlim} with $k\sim 0.3$. These limitations guarantee us applicability of first orders of perturbation theory for description of model (including the expressions of masses and couplings via  parameters of Lagrangian) and small value of quantum (loop) corrections.

\bu \ In  other models deviation of the triple Higgs coupling from its SM value can be stronger than that in 2HDM at moderate values of parameters, see  \cite{othermodhhh}. In the particular case of  the nMSSM (2HDM +Higgs singlet) values $\chi_{111}$ can range from -5 to 20 \cite{hhhMSSMn-1}, \cite{hhhMSSMn-2}.

\bu \  If the  mass $M_2$ of the
heavier Higgs boson $h_2$ lies  within the interval $(250\div 400)$~GeV and $|\chi^t_2|>1$  (in the SM-like scenario for $h_1$), the following interesting phenomenon takes place.
The boson $h_2$ becomes relatively narrow and the cross section of gluon fusion $gg\to h_2$ can be  larger than that for the would-be SM Higgs boson with mass $M_2$. The process $gg\to h_2\to h_1h_1$ can be seen as a resonant production of the $h_1h_1$ pair. In principle, it allows us to discover the mentioned $h_2$ at LHC (see examples in \cite{hhhviol1}, \cite{H+H-h_1}, \cite{h2hh} for special sets of parameters).\\

\section*{Acknowledgements}

The discussions with F. Boudjema, I.Ivanov, K.~Kanishev, M.~Krawczyk, P.~Osland, M.~Vysotsky were useful.

This work was supported in part by grants RFBR 15-02-05868,
NCN OPUS 2012/05/B/ST2/03306 (2012-2016) and HARMONIA project under contract
  UMO-2015/18/M/ST2/00518 (2016-2019).

\end{document}